# Scientific Data Management in the Coming Decade

Jim Gray, Microsoft
David T. Liu, Berkeley
Maria Nieto-Santisteban & Alexander S. Szalay, Johns Hopkins University
David DeWitt, Wisconsin
Gerd Heber, Cornell

January 2005

Technical Report
MSR-TR-2005-10

# Scientific Data Management in the Coming Decade

Jim Gray, Microsoft
David T. Liu, Berkeley
Maria Nieto-Santisteban & Alexander S. Szalay, Johns Hopkins University
David DeWitt, Wisconsin
Gerd Heber, Cornell

January 2005

## Data-intensive science – a new paradigm

Scientific instruments and computer simulations are creating vast data stores that require new scientific methods to analyze and organize the data. Data volumes are approximately doubling each year. Since these new instruments have extraordinary precision, the data quality is also rapidly improving. Analyzing this data to find the subtle effects missed by previous studies requires algorithms that can simultaneously deal with huge datasets and that can find very subtle effects – finding both needles in the haystack and finding very small haystacks that were undetected in previous measurements.

The raw instrument and simulation data is processed by pipelines that produce standard data products. In the NASA terminology[1], the raw *Level 0* data is calibrated and rectified to *Level 1* datasets that are combined with other data to make derived *Level 2* datasets. Most analysis happens on these *Level 2* datasets with drill down to *Level 1* data when anomalies are investigated.

We believe that most new science happens when the data is examined in new ways. So our focus here is on data exploration, interactive data analysis, and integration of *Level 2* datasets.

Data analysis tools have not kept pace with our ability to capture and store data. Many scientists envy the pen-and-paper days when all their data used to fit in a notebook and analysis was done with a slide-rule. Things were simpler then; one could focus on the science rather than needing to be an information-technology-professional with expertise in arcane computer data analysis tools.

The largest data analysis gap is in this man-machine interface. How can we put the scientist back in control of his data? How can we build analysis tools that are intuitive and that augment the scientist's intellect rather than adding to the intellectual burden with a forest of arcane user tools? The real challenge is building this *smart notebook* that unlocks the data and makes it easy to capture, organize, analyze, visualize, and publish.

This article is about the data and data analysis layer within such a smart notebook. We argue that *the smart notebook* will access data presented by science centers that will provide the community with analysis tools and computational resources to explore huge data archives.

## New data-analysis methods

The demand for tools and computational resources to perform scientific data-analysis is rising even faster than data volumes. This is a consequence of three phenomena: (1) More sophisticated algorithms consume more instructions to analyze each byte. (2) Many analysis algorithms are super-linear, often needing $N^2$ or $N^3$ time to process $N$ data points. And (3) IO bandwidth has not kept pace with storage capacity. In the last decade, while capacity has grown more than 100-fold, storage bandwidth has improved only about 10-fold.

These three trends: algorithmic intensity, nonlinearity, and bandwidth-limits mean that the analysis is taking longer and longer. To ameliorate these problems, scientists will need better analysis algorithms that can handle extremely large datasets with approximate algorithms (ones with near-linear execution time) and they will need parallel algorithms that can apply many processors and many disks to the problem to meet cpu-density and bandwidth-density demands.

## Science centers

These peta-scale datasets required a new work style. Today the typical scientist copies files to a local server and operates on the datasets using his own resources. Increasingly, the datasets are so large, and the application programs are so complex, that it is much more economical to move the end-user's programs to the data and only communicate questions and answers rather than moving the source data and its applications to the user's local system.

Science data centers that provide access to both the data and the applications that analyze the data are emerging as service stations for one or another scientific domain. Each of these science centers curates one or more massive datasets, curates the applications that provide access to that dataset, and supports a staff that understands the data and indeed is constantly adding to and improving the dataset. One can see this with the SDSS at Fermilab, BaBar at SLAC, BIRN at SDSC, with Entrez-PubMed-GenBank at NCBI, and with many other datasets across other disciplines. These centers federate with others. For example BaBar has about 25 peer sites and CERN LHC expects to have many Tier1 peer sites. NCBI has several peers, and SDSS is part of the International Virtual Observatory.

---

[1] Committee on Data Management, Archiving, and Computing (CODMAC) Data Level Definitions
http://science.hq.nasa.gov/research/earth_science_formats.html

The new work style in these scientific domains is to send questions to applications running at a data center and get back answers, rather than to bulk-copy raw data from the archive to your local server for further analysis. Indeed, there is an emerging trend to store a *personal workspace* (a *MyDB*) at the data center and deposit answers there. This minimizes data movement and allows collaboration among a group of scientists doing joint analysis. These personal workspaces are also a vehicle for data analysis groups to collaborate. Longer term, personal workspaces at the data center could become a vehicle for data publication – posting both the scientific results of an experiment or investigation along with the programs used to generate them in public read-only databases.

Many scientists will prefer doing much of their analysis at data centers because it will save them having to manage local data and computer farms. Some scientists may bring the small data extracts "home" for local processing, analysis and visualization – but it will be possible to do all the analysis at the data center using the personal workspace.

When a scientist wants to correlate data from two different data centers, then there is no option but to move part of the data from one place to another. If this is common, the two data centers will likely federate with one another to provide mutual data backup since the data traffic will justify making the copy.

Peta-scale data sets will require 1000-10,000 disks and thousands of compute nodes. At any one time some of the disks and some of the nodes will be broken. Such systems have to have a mechanism in place to protect against data loss, and provide availability even with a less than full configuration — a self-healing system is required. Replicating the data in science centers at different geographic locations is implied in the discussion above. Geographic replication provides both data availability and protects against data loss. Within a data center one can combine redundancy with a clever partitioning strategy to protect against failure at the disk controller or server level. While storing the data twice for redundancy, one can use different organizations (e.g. partition by space in one, and by time in the other) to optimize system performance. Failed data can be automatically recovered from the redundant copies with no interruption to database access, much as RAID5 disk arrays do today.

All these scenarios postulate easy data access, interchange and integration. Data must be self-describing in order to allow this. This self-description, or metadata, is central to all these scenarios; it enables generic tools to understand the data, and it enables people to understand the data.

## Metadata enables data access

Metadata is the descriptive information about data that explains the measured attributes, their names, units, precision, accuracy, data layout and ideally a great deal more. Most importantly, metadata includes the data lineage that describes how the data was measured, acquired or computed.

If the data is to be analyzed by generic tools, the tools need to "understand" the data. You cannot just present a bundle-of-bytes to a tool and expect the tool to intuit where the data values are and what they mean. The tool will want to know the metadata.

To take a simple example, given a file, you cannot say much about it – it could be anything. If I tell you it is a JPEG, you know it is a bitmap in http://www.jpeg.org/ format. JPEG files start with a header that describes the file layout, and often tells the camera, timestamp, and program that generated the picture. Many programs know how to read JPEG files and also produce new JPEG files that include metadata describing how the new image was produced. MP3 music files and PDF document files have similar roles – each is in a standard format, each carries some metadata, and each has an application suite to process and generate that file class.

If scientists are to read data collected by others, then the data must be carefully documented and must be published in forms that allow easy access and automated manipulation. In an ideal world there would be powerful tools that make it easy to capture, organize, analyze, visualize, and publish data. The tools would do data mining and machine learning on the data, and would make it easy to script workflows that analyze the data. Good metadata for the inputs is essential to make these tools automatic. Preserving and augmenting this metadata as part of the processing (data lineage) will be a key benefit of the next-generation tools.

All the derived data that the scientist produces must also be carefully documented and published in forms that allow easy access. Ideally much of this metadata would be automatically generated and managed as part of the workflow, reducing the scientist's intellectual burden.

## Semantic convergence: numbers to objects

Much science data is in the form of numeric arrays generated by instruments and simulations. Simple and convenient data models have evolved to represent arrays and relationships among them. These data models can also represent data lineage and other metadata by including narrative text, data definitions, and data tables within the file. HDF[2], NetCDF[3] and FITS[4] are good examples of such standards. They each include a library that encapsulates the files and provides a platform-independent way to read sub-arrays and to create or update files. Each standard allows easy data interchange among scientists. Generic tools that analyze and visualize these higher-level file formats are built atop each of these standards.

While the commercial world has standardized on the relational data model and SQL, no single standard or tool has critical mass in the scientific community. There are many parallel and competing efforts to build these tool suites – at least one per discipline. Data interchange outside each group is problematic. In the next decade, as data interchange among scientific disciplines becomes increasingly important, a common HDF-like format and package for all the sciences will likely emerge.

Definitions of common terminology (units and measurements) are emerging within each discipline. We are most familiar with the Universal Content Descriptors (UCD[5]) of the Astronomy community that define about a thousand core astrophysics units, measurements, and concepts. Almost every discipline has an analogous ontology (a.k.a., *controlled vocabulary*) effort. These efforts will likely start to converge over the next decade – probably as part of the converged format standard. This will greatly facilitate tool-building and tools since an agreement on these concepts can help guide analysis tool designs.

In addition to standardization, computer-usable ontologies will help build the Semantic Web: applications will be semantically compatible beyond the mere syntactic compatibility that current-generation of Web services offer with type matching interfaces. However, it will take some time before high-performance general-purpose *ontology engines* will be available and integrated with data analysis tools.

Database users on the other hand are well positioned to prototype such applications: a database schema, though not a complete ontology in itself, can be a rich ontology extract. SQL can be used to implement a rudimentary *semantic algebra*. The XML integration in modern Database Management Systems (DBMS) opens the door for existing standards like RDF and OWL.

Visualization or better *visual exploration* is a prime example of an application where success is determined by the ability to map a question formulated in the conceptual framework of the domain ontology onto the querying capabilities of a (meta-) data analysis backend. For the time being, a hybrid of SQL and XQuery is the only language suitable to serve as the target assembly language in this translation process.

## Metadata enables data independence

The separation of data and programs is artificial – one cannot see the data without using a program and most programs are data driven. So, it is paradoxical that the data management community has worked for 40 years to achieve something called *data independence* – a clear separation of programs from data. Database systems provide two forms of data independence termed *physical data independence* and *logical data independence*.

*Physical data independence* comes in many different forms. However, in all cases the goal is to be able to change the underlying physical data organization without breaking any application programs that depend on the old data format. One example of physical data independence is the ability of a database system to partition the rows of a table across multiple disks and/or multiple nodes of a cluster without requiring that any application programs be modified. The mapping of the fields of each row of a relational table to different disks is another important example of physical data independence. While a database system might choose to map each row to a contiguous storage container (e.g. a record) on a single disk page, it might also choose to store large, possibly infrequently referenced attributes of a table corresponding to large text objects, JPEG images, or multidimensional arrays in separate storage containers on different disk pages and/or different storage volumes in order to maximize the overall performance of the system. Again, such physical storage optimizations are implemented to be completely transparent to application programs except, perhaps, for a change in their performance. In the scientific domain the analogy would be that you could take a working application program that uses a C struct to describe its data records on disk and change the physical layout of the records without having to rewrite or even recompile the application program (or any of the other application programs that access the same data). By allowing such techniques, physical data independence allows performance improvements by reorganizing data for

---

[2] http://hdf.ncsa.uiuc.edu/HDF5/
[3] http://my.unidata.ucar.edu/content/software/netcdf/
[4] http://fits.gsfc.nasa.gov/
[5] http://vizier.u-strasbg.fr/doc/UCD.htx

parallelism–at little or no extra effort on the part of scientists.

Modern database systems also provide *logical data independence* that insulates programs from changes to the logical database design – allowing designers to add or delete relationships and to add information to the database. While physical data independence is used to hide changes in the physical data organizations, logical data independence hides changes in the logical organization of the data. Logical data independence is typically supported using *views*. A view defines a virtual table that is specified using a SQL query over one or more base tables and/or other views. Views serve many purposes including increased security (by hiding attributes from applications and/or users without a legitimate need for access) and enhanced performance (by materializing views defined by complex SQL queries over very large input tables). But views are primarily used to allow old programs to operate correctly even as the underlying database is reorganized and redesigned. For example, consider a program whose correct operation depends on some table T that a database administrator wants to reorganize by dividing vertically into two pieces stored in tables T' and T". To preserve applications that depend on T, the database administrator can then define a view over T' and T" corresponding to the original definition of table T, allowing old programs to continue to operate correctly.

In addition, data evolves. Systems evolve from EBCDIC to ASCII to Unicode, from proprietary-float to IEEE-float, from marks to euros, and from 8-character ASCII names to 1,000 character Unicode names. It is important to be able to make these changes without breaking the millions of lines of existing programs that want to see the data in the old way. Views are used to solve these problems by dynamically translating data to the appropriate formats (converting among character and number representations, converting among 6-digit and 9-digit postal codes, converting between long-and-short names, and hiding new information from old programs.) The pain of the Y2K (converting from 2-character to 4-character years) taught most organizations the importance of data independence.

Database systems use a *schema* to implement both logical and physical data independence. The schema for a database holds all metadata including table and view definitions as well as information on what indices exist and how tables are mapped to storage volumes (and nodes in a parallel database environment). Separating the data and the metadata from the programs that manipulate the data is crucial to data independence. Otherwise, it is essentially impossible for other programs to find the metadata which, in turn, makes it essentially impossible for multiple programs to share a common database. Object-oriented programming concepts have refined the separation of programs and data. Data classes encapsulated with methods provide data independence and make it much easier to evolve the data without perturbing programs. So, these ideas are still evolving.

But the key point of this section is that an explicit and standard data access layer with precise metadata and explicit data access is essential for data independence.

## Set-oriented data access gives parallelism

As mentioned earlier, scientists often start with numeric data arrays from their instruments or simulations. Often, these arrays are accompanied by tabular data describing the experimental setup, simulation parameters, or environmental conditions. The data are also accompanied by documents that explain the data.

Many operations take these arrays and produce new arrays, but eventually, the arrays undergo *feature extraction* to produce *objects* that are the basis for further analysis. For example, raw astronomy data is converted to object catalogs of stars and galaxies. Stream-gauge measurements are converted to stream-flow and water-quality time-series data, serum-mass-spectrograms are converted to records describing peptide and protein concentrations, and raw high-energy physics data are converted to events.

Most scientific studies involve exploring and data mining these object-oriented tabular datasets. The scientific file-formats of HDF, NetCDF, and FITS can represent tabular data but they provide minimal tools for searching and analyzing tabular data. Their main focus is getting the tables and sub-arrays into your Fortran/C/Java/Python address space where you can manipulate the data using the programming language.

This Fortran/C/Java/Python file-at-a-time procedural data analysis is nearing the breaking point. The data avalanche is creating billions of files and trillions of events. The file-oriented approach postulates that files are organized into directories. The directories relate all data from some instrument or some month or some region or some laboratory. As things evolve, the directories become hierarchical. In this model, data analysis proceeds by searching all the relevant files – opening each file, extracting the relevant data and then moving onto the next file. When all the relevant data has been gathered in memory (or in intermediate files) the program can begin its analysis. Performing this *filter-then-analyze,* data analysis on large datasets with conventional procedural tools runs slower and slower as data volumes increase. Usually, they use only one-cpu-at-a-time; one-disk-at-a-time and they do a brute-force search of the data. Scientists need a way (1) to use intelligent indices and data organizations to subset the search, (2) to use parallel processing and data access to search huge datasets within seconds, and (3) to have powerful analysis tools that they can apply to the subset of data being analyzed.

One approach to this is to use the MPI (Message Passing Interface) parallel programming environment to write procedural programs that stream files across a processor array – each node of the array exploring one part of the hierarchy. This is adequate for highly-regular array processing tasks, but it seems too daunting for ad-hoc analysis of tabular data. MPI and the various array file formats lack indexing methods other than partitioned sequential scan. MPI itself lacks any notion of metadata beyond file names.

As file systems grow to petabyte-scale archives with billions of files, the science community must create a synthesis of database systems and file systems. At a minimum, the file hierarchy will be replaced with a database that catalogs the attributes and lineage of each file. Set-oriented file processing will make file names increasingly irrelevant – analysis will be applied to "all data with these attributes" rather than working on a list of file/directory names or name patterns. Indeed, the files themselves may become irrelevant (they are just containers for data.) One can see a harbinger of this idea in the Map-Reduce approach pioneered by Google[6]. From our perspective, the key aspect of Google Map-Reduce is that it applies thousands of processors and disks to explore large datasets in parallel. That system has a very simple data model appropriate for the Google processing, but we imagine it could evolve over the next decade to be quite general.

The database community has provided automatic query processing along with CPU and IO parallelism for over two decades. Indeed, this automatic parallelism allows large corporations to mine 100-Terabyte datasets today using 1000 processor clusters. We believe that many of those techniques apply to scientific datasets[7].

## Other useful database features

Database systems are also approaching the peta-scale data management problem driven largely by the need to manage huge information stores for the commercial and governmental sectors. They hide the file concept and deal with data collections. They can federate many different sources letting the program view them all as a single data collection. They also let the program pivot on any data attributes.

Database systems provide very powerful data definition tools to specify the abstract data formats and also specify how the data is organized. They routinely allow the data to be replicated so that it can be organized in several ways (by time, by space, by other attributes). These techniques have evolved from mere indices to materialized views that can combine data from many sources.

Database systems provide powerful associative search (search by value rather than by location) and provide automatic parallel access and execution essential to peta-scale data analysis. They provide non-procedural and parallel data search to quickly find data subsets, and a many tools to automate data design and management.

In addition, data analysis using data cubes has made huge advances, and now efforts are focused on integrating machine learning algorithms that infer trends, do data clustering, and detect anomalies. All these tools are aimed at making it easy to analyze commercial data, but they are equally applicable to scientific data analysis.

## Ending the impedance mismatch

Conventional tabular database systems are adequate for analyzing objects (galaxies, spectra, proteins, events, etc.). But even there, the support for time-sequence, spatial, text and other data types is often awkward. Database systems have not traditionally supported science's core data type: the N-dimensional array. Arrays have had to masquerade as blobs (binary large objects) in most systems. This collection of problems is generally called the *impedance mismatch* – meaning the mismatch between the programming model and the database capabilities. The impedance mismatch has made it difficult to map many science applications into conventional tabular database systems.

But, database systems are changing. They are being integrated with programming languages so that they can support object-oriented databases. This new generation of object relational database systems treats any data type (be it a native float, an array, a string, or a compound object like an XML or HTML document) as an encapsulated type that can be stored as a value in a field of a record. Actually, these systems allow the values to be either stored directly in the record (embedded) or to be pointed to by the record (linked). This linking-embedding object model nicely accommodates the integration of database systems and file systems – files are treated as linked-objects. Queries can read and write these extended types using the same techniques they use on native types.

---

[6] "MapReduce: Simplified Data Processing on Large Clusters," J. Dean, S. Ghemawat, ACM OSDI, Dec. 2004.

[7] "Parallel Database Systems: the Future of High Performance Database Systems", D. DeWitt, J. Gray, CACM, Vol. 35, No. 6, June 1992.

Indeed we expect HDF and other file formats to be added as types to most database systems.

Once you can put your types and your programs inside the database you get the parallelism, non-procedural query, and data independence advantages of traditional database systems. We believe this database, file system, and programming language integration will be the key to managing and accessing peta-scale data management systems in the future.

## What's wrong with files?

Everything builds from files as a base. HDF uses files. Database systems use files. But, file systems have no metadata beyond a hierarchical directory structure and file names. They encourage a do-it-yourself- data-model that will not benefit from the growing suite of data analysis tools. They encourage do-it-yourself-access-methods that will not do parallel, associative, temporal, or spatial search. They also lack a high-level query language. Lastly, most file systems can manage millions of files, but by the time a file system can deal with billions of files, it has become a database system.

As you can see, we take an ecumenical view of what a database is. We see NetCDF, HDF, FITS, and Google Map-Reduce as nascent database systems (others might think of them as file systems). They have a schema language (metadata) to define the metadata. They have a few indexing strategies, and a simple data manipulation language. They have the start of non-procedural and parallel programming. And, they have a collection of tools to create, access, search, and visualize the data. So, in our view they are simple database systems.

## Why scientists don't use databases today

Traditional database systems have lagged in supporting core scientific data types but they have a few things scientists desperately need for their data analysis: non-procedural query analysis, automatic parallelism, and sophisticated tools for associative, temporal, and spatial search.

If one takes the controversial view that HDF, NetCDF, FITS, and Root are nascent database systems that provide metadata and portability but lack non-procedural query analysis, automatic parallelism, and sophisticated indexing, then one can see a fairly clear path that integrates these communities.

Some scientists use databases for some of their work, but as a general rule, most scientists do not. Why? Why are tabular databases so successful in commercial applications and such a flop in most scientific applications? Scientific colleagues give one or more of the following answers when asked why they do not use databases to manage their data:
- We don't see any benefit in them. The cost of learning the tools (data definition and data loading, and query) doesn't seem worth it.
- They do not offer good visualization/plotting tools.
- I can handle my data volumes with my programming language.
- They do not support our data types (arrays, spatial, text, etc.).
- They do not support our access patterns (spatial, temporal, etc.).
- We tried them but they were too slow.
- We tried them but once we loaded our data we could no longer manipulate the data using our standard application programs.
- They require an expensive guru (database administrator) to use.

All these answers are based on experience and considerable investment. Often the experience was with older systems (a 1990 vintage database system) or with a young system (an early object-oriented database or an early version of Postgres or MySQL.) Nonetheless, there is considerable evidence that databases have to improve a lot before they are worth a second look.

## Why things are different now

The thing that forces a second look now is that the file-ftp *modus operandi* just will not work for peta-scale datasets. Some new way of managing and accessing information is needed. We argued that metadata is the key to this and that a non-procedural data manipulation language combined with data indexing is essential to being able to search and analyze the data.

There is a convergence of file systems, database systems, and programming languages. Extensible database systems use object-oriented techniques from programming languages to allow you to define complex objects as native database types. Files (or extended files like HDF) then become part of the database and benefit from the parallel search and metadata management. It seems very likely that these nascent database systems will be integrated with the main-line database systems in the next decade or that some new species of metadata driven analysis and workflow system will supplant both traditional databases and the science-specific file formats and their tool suites.

## Some hints of success

There are early signs that this is a good approach. One of us has shown that the doing analysis atop a database system is vastly simpler and runs much faster than the corresponding file-oriented approach[8]. The speedup is due to better indexing and parallelism.

We have also had considerable success in adding user defined functions and stored procedures to astronomy databases. The MyDB and CasJobs work for the Sloan Digital Sky Survey give a good example of moving-programs-to-the-database[9].

The BaBar experiments at SLAC manage a petabyte store of event data. The system uses a combination of Oracle to manage some of the file archive and also a physics-specific data analysis system called Root for data analysis[10].

Adaptive Finite Element simulations spend considerable time and programming effort on input, output, and checkpointing. We (Heber) use a database to represent large Finite Element models. The initial model is represented in the database and each checkpoint and analysis step is written to the database. Using a database allows queries to define more sophisticated mesh partitions and allows concurrent indexed access to the simulation data for visualization and computational steering. Commercial Finite Element packages each use a proprietary form of a "database". They are, however, limited in scope, functionality, and scalability, and are typically buried inside the particular application stack. Each worker in the MPI job gets its partition from the database (as a query) and dumps its progress to the database. These dumps are two to four orders of magnitude larger than the input mesh and represent a performance challenge in both traditional and database environments. The database approach has the added benefit that visualization tools can watch and steer the computation by reading and writing the database. Finally, while we have focused on the ability of databases to simplify and speedup the production of raw simulation data, we cannot understate its core competency: providing declarative data analysis interfaces. It is with these tools that scientists spend most of their time. We hope to apply similar concepts to some turbulence studies being done at Johns Hopkins.

---

[8] "When Database Systems Meet the Grid," M. Nieto Santisteban et. al., CIDR, 2005,
http://www-db.cs.wisc.edu/cidr/papers/P13.pdf

[9] "Batch is back: CasJobs serving multi-TB data on the Web," W. O'Mullane, et. al, in preparation.

[10] "Lessons Learned from Managing a Petabyte," J. Becla and D. L. Wang, CIDR, 2005,
http://www-db.cs.wisc.edu/cidr/papers/P06.pdf

## Summary

Science centers that curate and serve science data are emerging around next-generation science instruments. The world-wide telescope, GenBank, and the BaBar collaborations are prototypes of this trend. One group of scientists is collecting the data and managing these archives. A larger group of scientists are exploring these archives the way previous generations explored their private data. Often the results of the analysis are fed back to the archive to add to the corpus.

Because data collection is now separated from data analysis, extensive metadata describing the data in standard terms is needed so people and programs can understand the data. Good metadata becomes central for data sharing among different disciplines and for data analysis and visualization tools.

There is a convergence of the nascent-databases (HDF, NetCDF, FITS,..) which focus primarily on the metadata issues and data interchange, and the traditional data management systems (SQL and others) that have focused on managing and analyzing very large datasets. The traditional systems have the virtues of automatic parallelism, indexing, and non-procedural access, but they need to embrace the data types of the science community and need to co-exist with data in file systems. We believe the emphasis on extending database systems by unifying databases with programming languages so that one can either embed or link new object types into the data management system will enable this synthesis.

Three technical advances will be crucial to scientific analysis: (1) extensive metadata and metadata standards that will make it easy to discover what data exists, make it easy for people and programs to understand the data, and make it easy to track data lineage; (2) great analysis tools that allow scientists to easily ask questions, and to easily understand and visualize the answers; and (3) set-oriented data parallelism access supported by new indexing schemes and new algorithms that allow us to interactively explore peta-scale datasets.

The goal is a *smart notebook* that empowers scientists to explore the world's data. Science data centers with computational resources to explore huge data archives will be central to enabling such notebooks. Because data is so large, and IO bandwidth is not keeping pace, moving code to data will be essential to performance. Consequently, science centers will remain the core vehicle and federations will likely be secondary. Science centers will provide both the archives and the institutional infrastructure to develop these peta-scale archives and the algorithms and tools to analyze them.